\documentclass[aps,prl,showpacs,twocolumn,floatfix]{revtex4}
\usepackage{graphicx}
\begin{document}

\draft
\title{Neutron drip line and the equation of state of nuclear matter}

\author{Kazuhiro Oyamatsu$^{1,2}$, Kei Iida$^{2,3}$, 
and Hiroyuki Koura$^{2,4}$}
\affiliation{$^1$Department of Human Informatics, Aichi Shukutoku
University, Nagakute, Nagakute-cho, Aichi-gun, Aichi 480-1197, Japan\\
$^2$RIKEN Nishina Center, RIKEN, Hirosawa, Wako, Saitama 351-0198, Japan\\
$^3$Department of Natural Science, Kochi University,
Akebono-cho, Kochi 780-8520, Japan\\
$^4$Advanced Science Research Center, Japan Atomic Energy Agency, Tokai, 
Ibaraki 319-1195, Japan
}

\date{\today}

\begin{abstract}
     We investigate how the neutron drip line is related to the density
dependence of the symmetry energy, by using a macroscopic nuclear model 
that allows us to calculate nuclear masses in a way dependent on the 
equation of state of asymmetric nuclear matter.  The neutron drip line
obtained from these masses is shown to appreciably shift to a neutron-rich 
side in a nuclear chart as the density derivative of the symmetry energy 
increases.  Such shift is clearly seen for light nuclei, a feature coming 
mainly from the surface property of neutron-rich nuclei.

\end{abstract}
\maketitle

%drip 

     Thanks to recent developments of radioactive ion beam facilities, one 
might be able to experimentally probe the stability of atomic nuclei against 
neutron or proton drip in a nuclear chart ranging from light to superheavy
nuclides.  The key quantities to study the neutron (proton) drip line are
the one- and two-neutron (proton) separation energies, $S_{n(p)}$ and 
$S_{2n(2p)}$, which correspond to an energy required to remove one and two 
neutrons (protons) from a nucleus in the ground state, respectively.  
Experimentally, the neutron drip line is marginally accessible only for light 
nuclei \cite{Notani}.  Even beyond the neutron drip line, however, nuclei can 
be present in dense neutral matter.
Nuclei in the crust of neutron stars are a typical example and are considered
to control the thermal and electric transport properties of matter in the
crust as well as the dynamics of superfluid neutron vortices, which are
relevant to the observed thermal and rotational evolution of neutron stars 
\cite{PR}.  We remark that the size and shape of nuclei are shown to be 
controlled by the equation of state (EOS) of asymmetric nuclear matter through 
the density dependence of the symmetry energy \cite{OI2}.  In this Letter, 
we will investigate how the density dependence of the symmetry energy in turn
affects the prediction of the neutron drip line.

%drip line from WB and other math formulas

    Theoretically, a Weizs{\" a}cker-Bethe mass formula, which is based on
a view of nuclei as incompressible spherical liquid drops of uniform 
density $n_0$, 
provides a standard behavior of the neutron drip line.  In this formula, the 
nuclear binding energy $E_B$ is written as function of mass number $A$ and 
charge number $Z$ (or neutron number $N$) in the form 
\begin{equation}
    -E_B=E_{\rm vol}+E_{\rm sym}+E_{\rm surf}+E_{\rm Coul},
\label{wb}
\end{equation}
where $E_{\rm vol}=a_{\rm vol}A$ is the volume energy, 
$E_{\rm sym}=a_{\rm sym}[(N-Z)/A]^2 A$ is the symmetry energy,
$E_{\rm surf}=a_{\rm surf}A^{2/3}$ is the surface energy, and 
$E_{\rm Coul}=a_{\rm Coul}Z^2/A^{1/3}$ is the Coulomb energy.  Then,
the one-neutron separation energy can be evaluated as
\begin{eqnarray}
   S_n &\approx& \left.\frac{\partial E_B}{\partial N}\right|_{Z}
\nonumber \\ &=&-a_{\rm vol}-a_{\rm sym}(1-4x^2)-\frac{2a_{\rm surf}}{3A^{1/3}}
     +\frac{a_{\rm Coul}Z^2}{3A^{4/3}},
 \label{sn}
\end{eqnarray}
where $x=Z/A$ is the proton fraction.  The condition $S_n=0$ gives a 
smoothed behavior of the neutron drip line.  The derived drip line is 
close to $x=0.3$, which is basically controlled by the competition 
between the volume and symmetry energy terms, except in the light region 
of the nuclear chart.  This may be a good starting point, but one needs
to go beyond Eq.\ (\ref{sn}) by taking into account nonnegligible deviation 
of the nuclear density from $n_0$.

%density dependence of symmetry energy %mass paper & strategy

     Recently, the density dependence of the symmetry energy 
attracts much attention because it is relevant to the isospin dependence of 
nuclear masses (e.g., Refs.\ \cite{DL,OI3}) and radii (e.g., Refs.\ 
\cite{OI1,BU,Warda}), dipole resonances (e.g., Refs.\ \cite{PDR,GDR}), and 
heavy-ion collisions involving neutron-rich nuclei (e.g., Refs.\ 
\cite{Shetty,LCK,Tsang}).  In predicting the neutron 
drip line, uncertainties in the density dependence of the symmetry energy 
need to be taken seriously.  The important parameter characterizing the
density dependence of the symmetry energy is a density symmetry coefficient 
$L$, which is defined as $L=3n_0(dS/dn)_{n=n_0}$ with the symmetry energy 
$S(n)$ dependent on the density $n$ of bulk nuclear matter.  Masses of 
extremely neutron-rich nuclei were calculated from a macroscopic nuclear
model and shown to have an appreciable dependence on $L$, which can 
be understood from the density and isospin dependence of the surface 
tension \cite{OI3}.  Here we address how this dependence affects
the neutron drip line on the nuclear chart.

%macroscopic nuclear model

     We begin with a macroscopic model of nuclei \cite{OI1}, which was 
constructed in such a way as to reproduce the known global properties of 
stable nuclei and can be used for describing the masses and radii of 
unstable nuclei in a manner that is dependent on the EOS of nuclear matter. 
This model can be summarized as follows:

\noindent
(i) We set the bulk energy per nucleon as
 \begin{eqnarray}
  w&=&\frac{3 \hbar^2 (3\pi^2)^{2/3}}{10m_n n}(n_n^{5/3}+n_p^{5/3})
   \nonumber \\ & & +(1-\alpha^2)v_s(n)/n+\alpha^2 v_n(n)/n,
\label{eos1}
\end{eqnarray}
where 
\begin{equation}
  v_s=a_1 n^2 +\frac{a_2 n^3}{1+a_3 n}
\label{vs}
\end{equation}
and
\begin{equation}
  v_n=b_1 n^2 +\frac{b_2 n^3}{1+b_3 n}
\label{vn}
\end{equation}
are the potential energy densities for symmetric nuclear matter and pure 
neutron matter, $n_n$ and $n_p$ are the neutron and proton number densities,
$n=n_n+n_p$, $\alpha=(n_n-n_p)/n$ is the neutron excess, and $m_n$ is the 
neutron mass.  A set of expressions (\ref{eos1})--(\ref{vn}) is one of the 
simplest that reduces to a usual expansion \cite{L}
\begin{equation}
    w=w_0+\frac{K_0}{18n_0^2}(n-n_0)^2+ \left[S_0+\frac{L}{3n_0}(n-n_0)
      \right]\alpha^2
\label{eos0}
\end{equation}
in the limit of $n\to n_0$ and $\alpha\to0$.  Here $w_0$ and $K_0$ are the 
saturation energy and the incompressibility of symmetric nuclear matter, and
$S_0=S(n=n_0)$.  In the incompressible limit, $w_0$ and $S_0$ correspond to
$a_{\rm vol}$ and $a_{\rm sym}$ in the mass formula (\ref{wb}), respectively.
We fix $b_3$, which controls the EOS of matter for large 
neutron excess and high density, at 1.58632 fm$^3$.  
This value was obtained by one of the authors \cite{O} in 
such a way as to reproduce the neutron matter energy of Friedman and 
Pandharipande \cite{FP}.  Change in this parameter would make no significant
difference in the determination of the other parameters and the final 
results for nuclear masses.

\noindent
(ii)  We write down the total energy of a nucleus of mass number $A$ and 
charge number $Z$ as a function of the density distributions $n_n({\bf r})$ 
and $n_p({\bf r})$ in the form
\begin{equation}
 E=E_b+E_g+E_C+Nm_n c^2+Zm_p c^2,
\label{e}
\end{equation}
where 
\begin{equation}
  E_b=\int d^3 r n({\bf r})w\left(n_n({\bf r}),n_p({\bf r})\right)
\label{eb}
\end{equation}
is the bulk energy,
\begin{equation}
  E_g=F_0 \int d^3 r |\nabla n({\bf r})|^2
\label{eg}
\end{equation}
is the gradient energy with adjustable constant $F_0$,
\begin{equation}
  E_C=\frac{e^2}{2}\int d^3 r \int  d^3 r' 
      \frac{n_p({\bf r})n_p({\bf r'})}{|{\bf r}-{\bf r'}|}
\label{ec}
\end{equation}
is the Coulomb energy, and $m_p$ is the proton mass.

\noindent
(iii) For simplicity we use the following parametrization for 
the nucleon distributions $n_i(r)$ $(i=n,p)$: 
\begin{equation}
  n_i(r)=\left\{ \begin{array}{lll}
  n_i^{\rm in}\left[1-\left(\displaystyle{\frac{r}{R_i}}\right)^{t_i}\right]^3,
         & \mbox{$r<R_i,$} \\
             \\
         0,
         & \mbox{$r\geq R_i,$}
 \end{array} \right.
\label{ni}
\end{equation}
where $r$ is the distance from the center of the nucleus.
This parametrization allows for the central density, half-density radius, and 
surface diffuseness for neutrons and protons separately.  

\noindent
(iv) In order to construct the nuclear model in such a way as to reproduce 
empirical masses and radii of stable nuclei, we first extremize the binding 
energy with respect to the particle distributions for fixed $A$, five EOS 
parameters, and $F_0$.  Next, for various sets of the incompressibility and 
the density symmetry coefficient, we obtained the remaining three EOS 
parameters and the gradient coefficient by fitting the calculated optimal 
values of charge number, mass excess, root-mean-square (rms) charge radius to 
empirical data for stable nuclei on the smoothed $\beta$ stability line 
\cite{O}.  In the range of the parameters $0<L<160$ MeV and 180 MeV 
$\le K_0 \le 360$ MeV, as long as $K_0 S_0/3 n_0 L \gtrsim 200$ MeV fm$^3$, 
we obtained a reasonable fitting to such data.  
As a result of this fitting, the parameters $n_0$, $w_0$, $S_0$, and $F_0$ 
are constrained as $n_0=0.14$--0.17 fm$^{-3}$, $w_0=-16\pm1$ MeV, 
$S_0=25$--40 MeV, and $F_0=66\pm6$ MeV fm$^5$.  The fitting gives rise to a 
relation nearly independent of $K_0$,
\begin{equation}
  S_0\approx B+CL,
  \label{linear}
\end{equation}
where $B\approx28$ MeV and $C\approx0.075$.

%results from macroscopic nuclear model (i)

\begin{figure}[t]
\includegraphics[width=8.5cm]{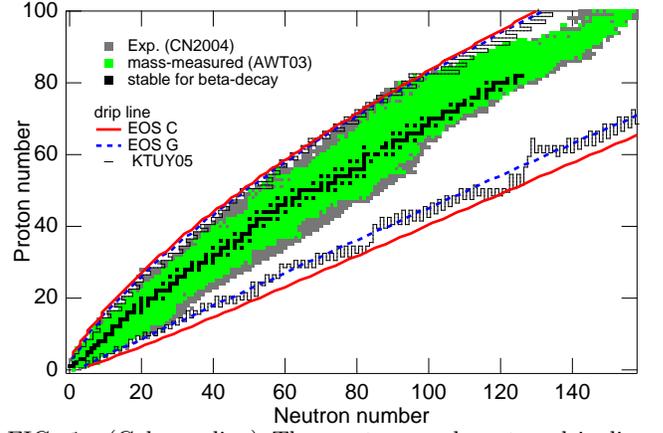} 
\vspace{-0.5cm}
\caption{\label{chart} (Color online)
The neutron and proton drip lines obtained from the EOS models C and G by
using the macroscopic nuclear model and from a contemporary mass formula 
\cite{KTUY}.  The regions filled with squares correspond to empirically 
known nuclides \cite{NC2004,AWT}.}
\end{figure}

    We proceed to obtain the neutron and proton drip lines from the 
macroscopic nuclear model.  For various sets of $L$ and $K_0$, we first 
evaluate the binding energy $E_B$ of nuclei in the ground state by 
minimizing the energy (\ref{e}) for fixed $N$ and $Z$.  We then draw the 
neutron (proton) drip line by identifying nuclides at neutron (proton)
drip with those neighboring to nuclides for which 
$S_{n}=E_B(Z,N)-E_B(Z,N-1)$ ($S_{p}=E_B(Z,N)-E_B(Z-1,N)$) and 
$S_{2n}=E_B(Z,N)-E_B(Z,N-2)$ ($S_{2p}=E_B(Z,N)-E_B(Z-2,N)$) are positive and 
beyond which at least one of them is negative.  The results obtained
from the two extreme EOS models denoted as EOS C ($L=146$ MeV and 
$K_0=360$ MeV) and EOS G ($L=5.7$ MeV and $K_0=180$ MeV) are shown in Fig.\ 1,
together with the empirically known nuclides \cite{NC2004,AWT} and the 
prediction from a contemporary mass formula \cite{KTUY}.  We remark that 
inclusion of the condition for $S_{2n}$ and $S_{2p}$ in addition to $S_n$ and 
$S_p$ in drawing 
the drip lines makes only a little difference in the case of the present 
model calculations, while being essential in the case of the prediction from 
the mass formula because of the Wigner, shell, and even-odd terms included 
therein.  We remark that the rms deviations of the calculated masses 
from the measured values \cite{AWT} are about 3 MeV, which is of order the
deviations obtained from a Weizs{\" a}cker-Bethe mass formula.

\begin{figure}[t]
\includegraphics[width=8.5cm]{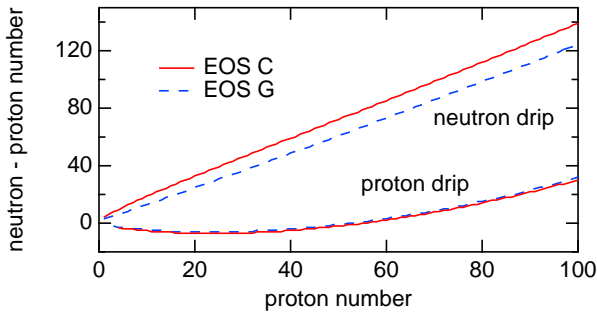}
\vspace{-0.5cm}
\caption{\label{I} (Color online)
$N-Z$ obtained for nuclides at neutron and proton drip from the EOS
models C and G.}
\end{figure}

    We find from Fig.\ 1 that the obtained neutron drip lines show an 
appreciable $L$ dependence, while the proton ones do not.  This is 
reasonable because nuclides at neutron (proton) drip are far away from (near) 
$N=Z$ (see Fig.\ 2).  The neutron drip line does shift to a neutron-rich 
side as $L$ increases, a feature that will be discussed later in terms of a
compressible liquid-drop model.  We remark that the EOS dependence of 
the obtained drip lines comes predominantly from $L$ because of negligible
$K_0$ dependence of the calculated masses \cite{OI3}.

\begin{figure}[t]
\includegraphics[width=8.5cm]{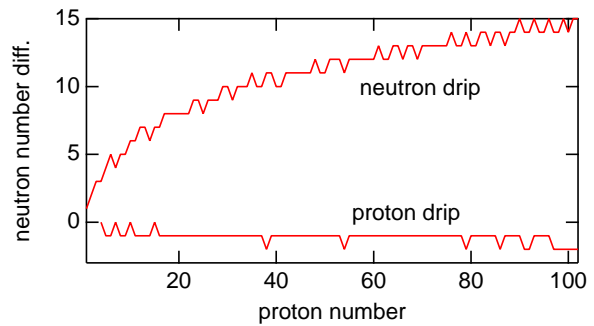}
\vspace{-0.5cm}
\caption{\label{Ndiff} (Color online)
Differences in the neutron number $N$ obtained for nuclides at neutron 
and proton drip between the calculations from the EOS models C and G.}
\end{figure}

    In order to see the $L$ dependence more clearly, we plot in 
Fig.\ 3 the difference in the neutron number of nuclides at neutron
and proton drip
between the calculations from the EOS models C and G.  The difference 
shows only a weak dependence on $Z$ both in the case of neutron and 
proton drip.  This indicates that the $L$ dependence can be seen more 
clearly for lighter nuclei.  In fact, the corresponding proton fraction 
of nuclides at neutron drip shows a stronger dependence on $L$
for lighter nuclei, as shown in Fig.\ 4.  This is advantageous because
heavier radioisotopes are more difficult to produce in experiments.

\begin{figure}[t]
\includegraphics[width=8.5cm]{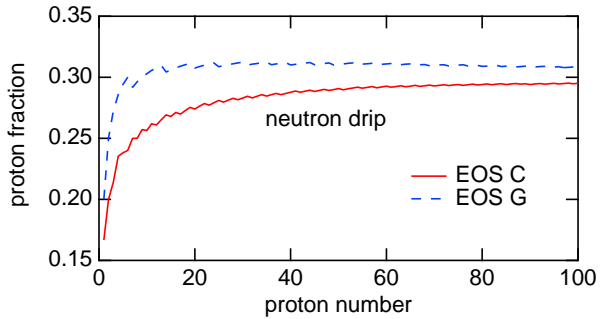}
\vspace{-0.5cm}
\caption{\label{xp} (Color online)
Proton fraction for nuclides at neutron drip 
obtained from the EOS models C and G.}
\end{figure}

%interpretation of the results by a compressible liquid-drop.

     The $L$ dependence of the neutron drip line as obtained above 
can be understood within the framework of a compressible liquid-drop 
model in which nuclei in equilibrium are allowed to have a density 
different from the saturation density $n_0$ of symmetric nuclear matter. 
By following a line of argument of Ref.\ \cite{OI3}, we first add the 
surface symmetry term, $a_{\rm ssym}A^{2/3}[(N-Z)/A]^2$, to the mass 
formula (\ref{wb}) based on an incompressible liquid-drop model.  
This surface symmetry term gives rise to additional contribution, 
\begin{equation}
 \delta S_n = -\frac{2a_{\rm ssym}}{3A^{1/3}}(1-2x)(1+4x), 
 \label{dSn}
\end{equation}
to the neutron separation 
energy (\ref{sn}).  Next, we consider the density-dependent surface
tension \cite{IO},
\begin{equation}
\sigma(n_{\rm in},\alpha_{\rm in})
=\sigma_0\left[1-C_{\rm sym}\alpha_{\rm in}^2
+\chi\left(\frac{n_{\rm in}-n_0}{n_0}\right)\right],
\label{sigma}
\end{equation}     
where $n_{\rm in}$ and $\alpha_{\rm in}$ are the density and neutron 
excess inside a liquid drop, $\sigma_0=\sigma(n_0,0)$, $C_{\rm sym}$ is 
the surface symmetry energy coefficient, and $\chi=(n_0/\sigma_0)
\partial\sigma/\partial n_{\rm in}|_{n_{\rm in}=n_0, \alpha_{\rm in}=0}$.
By taking a limit of vanishing compressibility, one obtains 
$4\pi\sigma_0 R^2=a_{\rm surf}A^{2/3}$ and 
$4\pi\sigma_0 C_{\rm sym} R^2=-a_{\rm ssym}A^{2/3}$ with the liquid-drop radius
$R$.  Typically, fitting to the empirical mass data yields 
$\sigma_0\approx1$ MeV fm$^{-2}$ and $C_{\rm sym}=1.5$--2.5.  As we shall 
see below, nonvanishing compressibility effectively introduces the $L$ 
dependence into the surface symmetry term through the parameter $\chi$ 
characterizing the density dependence of the surface tension.  The value
of $\chi$ is poorly known, but likely to be positive \cite{OI3}.  For example, 
$\chi=4/3$ for the Fermi gas model.

     If one ignores Coulomb and surface corrections, the equilibrium density 
and energy per nucleon of a liquid-drop, $n_s$ and $w_s$, can be evaluated 
from Eq.\ (\ref{eos0}) as
\begin{equation}
  w_s=w_0+S_0 \alpha^2
\label{ws}
\end{equation}
and
\begin{equation}
  n_s=n_0-\frac{3 n_0 L}{K_0}\alpha^2.
\label{ns}
\end{equation}
Strictly speaking, expressions (\ref{ws}) and (\ref{ns}) are applicable only 
for nearly symmetric nuclear matter.  We nevertheless use these expressions
for the purpose of characterizing the liquid-drop properties because the 
typical value of $\alpha$ along the neutron drip line is of order 0.35--0.4, 
considerably smaller than unity.  By substituting $n_s$, Eq.\ (\ref{ns}), 
into $n_{\rm in}$ in Eq.\ (\ref{sigma}), we thus obtain
\begin{equation}
\sigma(n_s,\alpha_{\rm in})
=\sigma_0\left[1-\left(C_{\rm sym}+\frac{3L\chi}{K_0}\right)
\alpha_{\rm in}^2\right].
\label{ndsigma}
\end{equation}     
This result allows one to identify $a_{\rm ssym}A^{2/3}$ with
$-4\pi \sigma_0 R^2 (C_{\rm sym}+3L\chi/K_0)$
and hence to conclude that with increasing $L$, $S_n$ increases 
through the surface symmetry contribution (\ref{dSn}) for $x<1/2$.

    This conclusion is consistent with the $L$ dependence of the 
neutron drip line shown in Fig.\ 1 because any positive corrections to
$S_n$ act to enhance the stability of nuclei against neutron emission.
It is important to note that the bulk symmetry term gives rise to
a negative contribution to $S_n$ for $x<1/2$, as shown in Eq.\ (\ref{sn}),
and that the parameter $a_{\rm sym}$ corresponds to the symmetry energy
coefficient $S_0$, which in turn is related to $L$ by the relation
(\ref{linear}) obtained from fitting to empirical masses 
and charge radii of stable nuclei.  Since one obtains a larger $a_{\rm sym}$ 
for larger $L$, the effect of $a_{\rm sym}$ tends to decrease $S_n$ with $L$
and hence to facilitate neutron drip.  However, this effect is relatively
small compared with the above-mentioned effect of $a_{\rm ssym}$.  This is 
consistent with the fact that the $L$ dependence is clearer for lighter 
nuclei.

%summary  pairing vs. smoothing

     In summary, we have investigated the influence of the density 
dependence of the symmetry energy on the drip lines by using a macroscopic 
nuclear model that depends explicitly on the EOS of nuclear matter.  
We find that an $L$ dependence appears appreciably in the neutron drip 
line and it is clearer for lighter nuclei, a feature coming mainly 
from the surface property through the density and neutron excess 
dependence of the surface tension.  We note that our calculations do not 
include even-odd or shell corrections.  The even-odd corrections would 
play an important role in $S_n$.  In fact, the magnitude of even-odd 
staggering in $S_n$ could be comparable to that of change in $S_n$ due 
to uncertainties in $L$.  This implies that some kind of smoothing would 
be required in deriving information about the EOS from future empirical 
data on neutron drip.  Shell corrections would further complicate such
derivation, but are intriguing in the context of magicity of nuclei 
close to the neutron drip line \cite{N16}.

\acknowledgments

      We are grateful to Dr.\ A. Kohama for useful discussion and 
late Prof.\ M. Uno for continuous encouragements.  Authors
K.O. and K.I. acknowledge the hospitality of the Yukawa Institute for 
Theoretical Physics during the workshop ``New Frontiers
in QCD 2010,'' where this work was initiated.  This work was supported 
in part by Grant-in-Aid for Scientific Research through Grant No.\ 
19740151, which was provided by the Ministry of Education, Culture, 
Sports, Science, and Technology of Japan.

\end{document}